# Distinct quantum anomalous Hall ground states induced by magnetic disorders

Chang Liu[1,2*], Yunbo Ou[1,*], Yang Feng[1], Gaoyuan Jiang[1], Weixiong Wu[1], Shaorui Li[1], Zijia Cheng[1], Ke He[1,3†], Xucun Ma[1,3], Qikun Xue[1,2,3], Yayu Wang[1,3†]

[1]*State Key Laboratory of Low Dimensional Quantum Physics, Department of Physics, Tsinghua University, Beijing 100084, P. R. China*

[2]*Beijing Academy of Quantum Information Sciences, Beijing 100193, P. R. China*

[3]*Frontier Science Center for Quantum Information, Beijing 100084, P. R. China*

* These authors contributed equally to this work.

[†] Emails: kehe@tsinghua.edu.cn; yayuwang@tsinghua.edu.cn

## ABSTRACT

**The quantum anomalous Hall (QAH) effect in magnetic topological insulator (TI) represents a new state of matter originated from the interplay between topology and magnetism. The defining characteristics of the QAH ground state are the quantized Hall resistivity ($\rho_{yx}$) and vanishing longitudinal resistivity ($\rho_{xx}$) in the absence of external magnetic field. A fundamental question concerning the QAH effect is whether it is merely a zero-magnetic-field quantum Hall (QH) effect, or if it can host unique quantum phases and phase transitions that are unavailable elsewhere. The most dramatic departure of the QAH systems from other QH systems lies in the strong magnetic disorders that induce spatially random magnetization. Because disorder and magnetism play pivotal roles in the phase diagram of two-dimensional electron systems, the high degree of magnetic disorders in QAH systems may create novel phases and quantum critical phenomena. In this work, we perform systematic transport studies of a series of magnetic TIs with varied strength of magnetic disorders. We find that the ground state of QAH effect can be categorized into two distinct classes: the QAH insulator and anomalous Hall (AH) insulator phases, as the zero-magnetic-field counterparts of the QH liquid and Hall insulator in the QH systems. In the low disorder limit of the QAH insulator regime, we observe a universal quantized longitudinal resistance $\rho_{xx} = h/e^2$ at the coercive field. In the AH insulator regime, we find that a magnetic field can drive it to the QAH insulator phase through a quantum critical point with distinct scaling behaviors from that in the QH phase transition. We propose that the transmission between chiral edge states at domain boundaries, tunable by disorder and magnetic fields, is the key for determining the QAH ground state.**

## I. INTRODUCTION

The discovery of QAH effect in magnetic TIs [1-3] represents a significant progress in the field of topological phase of matters. The QAH effect is generally regarded as a zero-magnetic-field QH effect, but a thorough examination reveals fundamental distinctions between them. Firstly, the QH effect requires the formation of discrete Landau levels generated by strong magnetic field, whereas the QAH effect relies on the exchange gap at the Dirac point arising from the interplay between topology and magnetic order [4-6]. The Dirac fermion's spin degree of freedom plays a central role in the QAH effect, thus its ground state is highly sensitive to the magnetic properties. Secondly, the charge carriers in the QH effect have high mobility due to the relatively weak potential disorder, whereas the QAH system is strongly disordered due to the random distribution of magnetic dopants, as demonstrated by a variety of scanning probe experiments [7-9]. The magnetic disorders in QAH samples cause strong spatial fluctuations of exchange gap size, making the system peculiar from the magnetic-disorder-free QH system [7]. Therefore, magnetic TI is highly unique for exploring disorder-related magnetic and topological quantum phenomena [10-17].

In the past few years, the QAH effect has been widely investigated in magnetic TIs with diverse experimental parameters, including the type and density of magnetic dopants, chemical compositions, film thicknesses, and device configurations [1-3,18-27]. However, it is still far from clear how the magnetic disorders affect the QAH ground states. Previous study on this direction mainly focused on the effect of on-site potential disorder on the transport properties [28], in analogy to the QH system, which leads to the finding of Anderson insulator phase in a QAH sample. Later spectroscopy experiments suggest that the magnetic disorder plays a more significant role on properties of magnetic TIs [8,29]. However, from transport perspective, the role of magnetic disorder has not been thoroughly investigated, mainly due to the experimental challenges in manipulating and detecting the influence of magnetic disorders. The debates regarding the chiral Majorana mode in QAH films coupled with a superconductor highlight the importance of understanding the role of strong magnetic disorders [30-33]. Therefore, it is interesting to explore if the QAH system can host any novel quantum phases and phase transitions that are unavailable elsewhere.

Before elucidating the ground states of the QAH effect, it is instructive to survey the global phase diagram of the QH effect, which has been established theoretically [34]. Based on the Chern-Simons theories, Kivelson, Lee & Zhang predicted that two distinct phases, namely the QH liquid and Hall insulator, exist in different disorder or magnetic field regimes and can be characterized by different temperature ($T$) dependent transport behaviors. For the ground state of a QH liquid, both the longitudinal conductivity $\sigma_{xx}$ and resistivity $\rho_{xx}$ become zero, whereas the Hall conductivity $\sigma_{xy}$ takes the quantized value $(e^2/h)s_{xy}$, where $e^2/h$ represents the quantum conductance and $s_{xy}$ is a particular rational fraction. For the Hall insulator phase, both $\sigma_{xx}$ and $\sigma_{xy}$ vanish at zero $T$, whereas $\rho_{xx}$ goes to infinity and $\rho_{yx}$ takes the classical value ~ $B/ne$ ($B$ denotes the magnetic field and $n$ is the carrier density). Later experiments not only confirmed the existence of the two phases and the universal quantum phase transition between them [35-37], but also discovered an unexpected quantized Hall insulator phase characterized by a quantized $\rho_{yx}$ rather than the classical value [38]. However, in the development of QAH effect so far, the zero-magnetic-field version of the QH liquid and Hall insulator has seldomly been investigated, especially in experiment.

## II. EXPERIMENTAL RESULTS AND DATA ANALYSIS

### A. Experimental methodology

Recently there have been more theoretical interests on the effect of magnetic disorder on the QAH effect [10-14], but a thorough experimental investigation has been lacking. The main reason lies in the fact that magnetic disorder is introduced by the random deposition of magnetic dopants during sample growth, probably with a Gaussian-like distribution [8,10], thus it is difficult to control the strength of magnetic disorder. Recently, we show that co-doping Cr and V magnetic elements during the epitaxial growth of TI thin films is an effective way to reduce the magnetic inhomogeneity [9,20]. The accurate control of Cr-V ratio leads to the achievement of QAH effect at much higher $T$ than that of the Cr and V singly doped magnetic TIs.

In this work, we employ a statistical methodology to investigate the effect of the degree of magnetic disorders on the QAH ground state. We have performed transport measurements on 82 magnetic TIs, with the results of 16 samples displayed in the main text. All the data presented in this work are taken when the gate voltage ($V_g$) is tuned to the charge neutral point (CNP) of each sample if not specified. All the samples were grown under parameters similar to that in previous reports for achieving the QAH effect [1,20,23,27]. The large quantity of samples, in conjunction with the co-doping method, allows us to cover a wide range with different strength of magnetic disorders. We first classify these samples by their zero-field $\rho_{xx}$ values ($\rho_{xx}^0$), which reflect the degree of magnetic disorder of a sample without the influence of external magnetic field. Because $\rho_{xx}$ takes all kinds of scattering processes into account, this criterion was commonly adopted in the studies of quantum phase transitions in different disordered electronic systems and AH effect in ferromagnets [34,36,37,39-41]. By thoroughly analyzing the magnetic field dependent $\rho_{xx}$ and $\rho_{yx}$ at varied $T$s and $V_g$s, we discover two novel quantum phases: the QAH insulator and AH insulator. We propose that the transmission between chiral edge states at domain boundaries, tunable by the strength of magnetic disorder and magnetic fields, is the key for determining the QAH ground state in magnetic TIs. Our results also shed new lights on some longstanding issues regarding the peculiar transport phenomena in QAH effect.

### B. Classification of the QAH ground state

We start by presenting the transport data of five representative magnetic TI samples (denoted as #S1 to #S5) with different strength of magnetic disorder. The lowest disordered sample #S1 was grown by the Cr-V co-doping method [20], whereas the other four samples were Cr singly doped magnetic TIs with relatively higher level of disorder [8]. The details of all the samples studied in this work are documented in the supplementary information.

Figures 1(a) to 1(e) display the $T$ dependences of $\rho_{xx}$ and $\rho_{yx}$ for the five samples, with the corresponding chemical formulas shown on top of the figures. All the data were measured at $V_g$ = 0 V, when the chemical potential lies in the charge neutral regime (see supplementary information Fig. S1 for details). From the perspective of Hall effect, all five samples display qualitatively similar behaviors charateristics of the QAH effect [1-3]. With the decrease of $T$, the coercive field ($H_C$) increases and the AH resistivity $\rho_{yx}^0$ (defined as the $\rho_{yx}$ at zero field) grows towards the quantum resistance $h/e^2$. The only quantitative difference in $\rho_{yx}^0$ lies in the exact level of quantization. Sample #S1 with the lowest degree of disorder (smallest $\rho_{xx}^0$ value) exhibits the best quantization at $T$ = 1.6 K, with $\rho_{yx}^0$ approaching 0.96 $h/e^2$. Whereas in the most disordered sample #S5 (largest $\rho_{xx}^0$ value), the $\rho_{yx}^0$ is only 0.56 $h/e^2$ at $T$ = 1.6 K.

From the perspective of $\rho_{xx}$, however, the five samples exhibit dramatic distinctions in the qualitative $T$ dependent behaviors. In sample #S1, $\rho_{xx}^0$ decreases from 0.9 $h/e^2$ at $T$ = 35 K to as low as 0.2 $h/e^2$ at $T$ = 1.6 K. In sample #S5, in the contrary, $\rho_{xx}^0$ increases from 2.5 $h/e^2$ to 12 $h/e^2$ as $T$ is lowered from 30 K to 1.6 K. To better visualize such distinct behaviors, Fig. 1(f) summarizes the $T$ dependent $\rho_{xx}^0$ for the five samples extracted from Figs. 1(a) to 1(e). From the slope of the $\rho_{xx}^0$ vs. $T$ curves, we clearly observe a phase transition from metallic to insulating behavior with the increase of disorder from #1 to #5. The observed trend is highly analogous to the classification of QH liquid and Hall insulator in conventional QH system [34-38]. Inspired by such analogy, we divide the five samples into two distinct ground states, dubbed as the QAH insulator (#S1, #S2, and #S3) and AH insulator (#S4 and #S5) phases, respectively, as the zero-magnetic-field counterparts of the QH liquid and Hall insulator phases. The different topological properties of the QAH insulator and the AH insulator are also reflected in the distinct behaviors of the Hall conductivity $\sigma_{xy}$, which is directly linked to the Chern number ($C$) in the form of $\sigma_{xy} = Ce^2/h$. In the good QAH insulator sample #S1, $\sigma_{xy}^0$ is quantized at 1.004 $e^2/h$, whereas in the typical AH insulator sample #S5, it is only as small as 0.004 $e^2/h$. More details about the behaviors of conductivity in the two phases are documented in the supplementary Figs. S2 and S3. The horizontal black broken line in Fig. 1(f) at $\rho_{xx}^0 \sim 2\ h/e^2$ roughly marks the boundary between the QAH insulator and AH insulator phases.

Another highly unexpected feature of the QAH ground state concerns the $\rho_{xx}$ value at $H_C$ ($\rho_{xx}^{Hc}$), in which the magnetic configuration is the most random and complex. The $T$ dependent $\rho_{xx}^{Hc}$ values for the five samples are shown in Fig. 1(g). In the most disordered sample #S5, $\rho_{xx}^{Hc}$ displays highly insulating behavior, reaching as high as 20 $h/e^2$ at $T$ = 1.6 K. With the decrease of disorder, the insulating behavior becomes weaker. In the lowest disordered sample #S1, interestingly, $\rho_{xx}^{Hc}$ retains the quantum resistance $h/e^2$ over a wide $T$ range. The appearance of $h/e^2$ implies that a topologically nontrivial mechanism dominates the transport behavior near the $H_C$.

An instructive way to summarize the evolution of transport behavior with disorder is to plot $\rho_{xx}^{Hc}$ as a function of $\rho_{xx}^0$ for different $T$s, as shown in Fig. 1(h). It clearly shows that for the lowest disordered sample #S1, $\rho_{xx}^{Hc}$ remains a constant at $h/e^2$ for the entire $T$ range. With the increase of disorder (samples #S2 and #S3), the $\rho_{xx}^{Hc}$ vs. $\rho_{xx}^0$ traces have a negative slope that becomes steeper as $\rho_{xx}^0$ approaches 2 $h/e^2$, which is the boundary between the QAH insulator and AH insulator phases. When the disorder is further increased (samples #S4 and #S5), the slope changes sign to positive. The correlation between $\rho_{xx}^{Hc}$ and $\rho_{xx}^0$ can be summarized by a power law $\rho_{xx}^{Hc} \sim (\rho_{xx}^0)^{\alpha}$, and the $\alpha$ values for different samples are listed in the figure. For $\alpha = 0$, the sample is in the low disorder limit of the QAH insulator phase with $\rho_{xx}^{Hc} \sim h/e^2$. For $\alpha = 1$, the sample is in the high disorder limit of the AH insulator phase when the peak near $H_C$ is so broad that $\rho_{xx}^{Hc} \sim \rho_{xx}^0$. In fact, the $\rho_{xx}^{Hc} \sim h/e^2$ behavior was present in some previous data on QAH samples with varied parameters [18,20,22,24,26,28,33], although without being noticed.

### C. More transport data on typical QAH insulator and AH insulator samples

To demonstrate the universality of the distinct QAH ground states, in Figs. 2(a) and 2(b) we display transport data on more magnetic TI samples in the typical QAH insulator and AH insulator phases. All the data shown here are measured at $T$ = 1.6 K. Figure 2(a) shows the transport data of six low-disorder Cr-V co-doped samples, where the $\rho_{xx}^0$ values are all smaller than 0.5 $h/e^2$ and $\rho_{yx}$ all show excellent quantization with $\rho_{yx}^0 > 0.93$ $h/e^2$. Remarkably, their $\rho_{xx}^{Hc}$ all lie very close to $h/e^2$ despite strong variations of $H_C$, strikingly similar to that shown in Fig. 1(a). Figure 2(b) displays the transport data of five high-disorder Cr-doped AH insulator samples with $\rho_{xx}^0 > 3$ $h/e^2$, where $\rho_{yx}$ shows the signature of QAH effect albeit a less perfect quantization. Here the $\rho_{xx}^{Hc}$ values are much larger than $h/e^2$ and display strong variations, also similar to that shown in Figs. 1(d) and 1(e). In Fig. 2(c), we summarize the relationship between $\rho_{xx}^{Hc}$ and $\rho_{xx}^{0}$ of all the 16 samples discussed in the main text measured at $T$ = 1.6 K (see Supplementary Fig. S4 for the summary of all 82 samples and the comparison with the previous results). In the large $\rho_{xx}^{0}$ regime with strong disorder, $\rho_{xx}^{Hc}$ decreases monotonically with the decrease of $\rho_{xx}^{0}$. When $\rho_{xx}^{0}$ is reduced to below 0.5 $h/e^2$, $\rho_{xx}^{Hc}$ gradually saturates at the $h/e^2$ level for the entire low disorder regime. The vertical broken line near $\rho_{xx}^{0}$ ~ 2 $h/e^2$ marks the approximate boundary between the QAH insulator and AH insulator phases, when the slope of the $\rho_{xx}^{0}$ vs. $T$ curve changes sign in Fig. 1(f).

To further strengthen the classification of phase, we extend the transport measurements to much lower $T$s in two representative QAH insulator and AH insulator samples [#S6 and #S13 in Fig. 2(c)]. The chemical formulas are presented on top of the figures. The upper panels of Figs. 3(a) and 3(b) display the magnetic field dependent $\rho_{yx}$ of the two samples at varied $T$s, both showing the characteristic square-shaped hysteresis loops. The lower panels of Figs. 3(a) and 3(b) display the $\rho_{xx}$ curves of the two samples, both showing two sharp peaks at $H_C$. As marked by the black broken line, the $\rho_{xx}^{Hc}$ value of sample #S6 stays very close to $h/e^2$ over the whole $T$ range, although $H_C$ increases significantly with decreasing $T$. In contrast, the $\rho_{xx}^{Hc}$ value of sample #S13 is much larger than $h/e^2$ and increases rapidly with the decrease of $T$. The blue curves in Figs. 3(c) and 3(d) summarize the values of $\rho_{yx}^{0}$ in the two samples, both showing the tendency towards quantization at low $T$. In contrast, $\rho_{xx}^{0}$ of the two samples (red curves) have totally opposite $T$ dependence. In sample #S6, it shows a metallic behavior and decreases to less than 0.02 $h/e^2$ at $T$ = 300 mK, whereas in sample #S13 it shows an insulating behavior and increases to more than 4.1 $h/e^2$ at $T$ = 50 mK (see Supplementary Figs. S5 and S6 for the $\sigma_{xy}$ and $\sigma_{xx}$ data). These low-$T$ results undoubtedly validate the classification of the QAH insulator and AH insulator ground states.

Although the magnetic properties of QAH samples also affect the transport behavior, they are not crucial to the classification of the ground states. In order to demonstrate this point, we analyze the relation between the magnetic properties and $\rho_{xx}^{0}$, as shown in Fig. S7 in the supplementary information. The absence of direct correlation between the classification of the QAH ground states and the magnetic properties, including the magnetic anisotropy, the domain energy, and the Curie temperature $T_c$, unambiguously demonstrates the dominant role of magnetic disorder on the transport properties.

### D. Magnetic-field-induced quantum phase transition

In addition to disorder, magnetic field also plays a crucial role in determining the ground state of QAH system. Figure 4(a) displays the magnetic field dependent $\rho_{xx}$ from $T$ = 100 mK to 800 mK for sample #13, which is an AH insulator in zero magnetic field. The blue arrow marks the direction of field sweep, which ensures that the initial state is always fully magnetized by strong out-of-plane magnetic field. The overall trend is that $\rho_{xx}$ increases with the decrease of magnetic field, and the crossing at 0.25 T indicates the existence of a quantum critical point (QCP). In Fig. 4(b) we extract the $\rho_{xx}$ value of different magnetic fields and replot them as a function of $T$. It clearly shows that the value of $\rho_{xx}$ vs. $T$ curve at the transition field ($H_T$) is nearly a constant around 2.6 $h/e^2$, which sets a phase boundary between the metallic (high field) and insulating (low field) ground states. This value is close to the phase boundary $\rho_{xx}^0$ ~ 2 $h/e^2$ obtained in Fig. 1(f). Therefore, increasing magnetic field drives a quantum phase transition from AH insulator to QAH insulator phase, in a similar manner as reducing disorder.

In conventional QH effect, a universal scaling law governs the behavior of both the liquid-insulator transition [36,37] and plateau-plateau transition [42-45]. To check the validity of the so-called "super-universality" in the QAH effect [34,46], following the criteria based on the crossing point in resistivity curves [47], we perform the same scaling analysis of the quantum phase transition between the QAH insulator and AH insulator phases. Figure 4(c) shows that all the $\rho_{xx}$ vs $(H\text{-}H_T)/T^{\kappa}$ curves, where $(H\text{-}H_T)/T^{\kappa}$ is the single parameter scaling function, collapse into a universal curve for a critical parameter $\kappa$ ~ 0.31±0.01. The critical $\rho_{xx}$ value at the QCP is found to be 2.6 $h/e^2$, which sets a phase boundary between the metallic (high field) and insulating (low field) ground states. The value of $\kappa$ ~ 0.31 is about 26% smaller than $\kappa$ ~ 0.42 in the QH phase transition [36,37]. Meanwhile, the $\rho_{xx}$ value at the QCP deviates significantly from the theoretical expectation of $h/e^2$ for the QH phase transition [34-37,48]. Consistent result is obtained in another sample, in which $\kappa$ and critical $\rho_{xx}$ are found to be 0.29 and 1.7 $h/e^2$ (see supplementary Fig. S9 for details). These results strongly suggest that the quantum critical behavior in the QAH system is different from that in conventional QH systems, probably due to the existence of strong magnetic disorders.

## E. Edge state transmission model

Because a rigorous theory about the QAH phase transition is still lacking, here we take a phenomenological approach to give an intuitive explanation for the classification of phases and the transition between them. The basic idea is similar to the phase incoherent network model invented to describe the quantized Hall insulator phase in QH systems [49], in which electron transport proceeds via a network of edge states surrounding QH puddles. As long as the puddle size is larger than the phase coherent length, $\rho_{yx}$ is always quantized regardless of the $\rho_{xx}$ behavior [50,51]. The most unique feature of QAH systems is the existence of magnetic domains, and the chiral edge states at domain boundaries [8]. We propose that the transmission between the edge states of neighboring domains plays the most crucial role in the quantum transport process. In a macroscopic QAH sample, magnetic disorders lead to spatially separated and randomly distributed magnetic domains [7-15]. Under such situation, charge transport is mainly determined by two quantities, the average distance $\lambda$ between neighboring domains and the decay length $\xi$ of the chiral edge state at domain boundaries.

Without loss of generality, we start from the simplest situation consisting of two domains. MFM measurements in Cr-V codoped and Cr singly doped TI samples show that the value of $\lambda$ in a good QAH insulator is estimated to be around 100 nm. Whereas in AH insulator samples with relatively high degree of magnetic disorder, its value can be up to 1 μm [20]. The value of $\xi$ is in the same range, from a few hundred nanometers to 1 μm, as revealed by the visualization of spatial conductivity on a magnetic TI by MIM measurements [52]. Figure 5(a) schematically illustrates the magnetic domain distribution in an AH insulator, in which $\lambda$ is larger than $\xi$. The transmission between chiral edge states is by quantum mechanical hopping across the barrier, similar to variable range hopping in strongly disordered systems, thus becomes insulating at low $T$. With the reduction of disorder strength or increase of magnetic field, the average size of magnetic domain grows so that the distance $\lambda$ between chiral edge states becomes smaller, as shown in Fig. 5(b). When $\lambda$ becomes comparable to $\xi$, the chiral edge states overlap substantially and their chemical potentials equilibrate. Consequently, the transmission has an ohmic contact behavior, which corresponds to the metallic $\rho_{xx}$ behaviors in the QAH insulator regime. The quantum phase transition from the AH insulator to QAH insulator, achieved either by reducing disorder or increasing magnetic field, can be explained naturally in this picture.

The universal $\rho_{xx}^{H_c} \sim h/e^2$ observed in the low disorder limit of the QAH insulator phase can also be understood in terms of edge state transmission. As shown in Fig. 5(c), at the $H_C$ of a QAH insulator, the edge states of two neighboring domains with opposite magnetization achieve chemical potential equilibration due to substantial wavefunction overlap. When perfect transmission occurs at the domain boundary, the chemical potential of the two upstream edge states get fully equalized, so that the two upper edges have the same potential. Following the Landauer-Buttiker formalism [53], the $\rho_{xx}$ at different edges obtain different quantized values, 2 $h/e^2$ on the lower edge and 0 on the upper edge [numbers in Fig. 5(c) indicate the chemical potential distribution on different sections of the edge states]. This effect has been directly demonstrated by transport measurement in a two-domain device fabricated by the tip of a magnetic force microscope [53]. The macroscopic samples studied in our experiment consist of large numbers of up and down magnetic domains at $H_C$, and the average of the two quantized values gives rise to the observed quantum resistance $h/e^2$. The universal $\rho_{xx}^{H_c} \sim h/e^2$ behavior is thus a manifestation of topological edge state transport that obeys the Landauer-Buttiker formalism in the low disorder limit of QAH insulator phase [3,21,22,24] (see Supplementary session J and K for details).

## III. DISCUSSIONS AND CONCLUSIONS

The proposed phenomenological picture not only explains the formation and evolution of the QAH insulator and AH insulator phases, but also sheds important new lights on the peculiar transport phenomena in magnetic TIs. For example, the QAH state is theoretically expected to show quantized $\rho_{yx}$ and vanishing $\rho_{xx}$ simultaneously due to the dissipationless edge states. However, numerous previous experiments show that $\rho_{yx}$ can be well quantized at zero magnetic field, but $\rho_{xx}$ is always finite (up to a few $h/e^2$) [1-3,18-28]. Several theoretical models have been proposed to explain this discrepancy [10,15,54], but there is no consensus yet. Our results provide a new physical picture to resolve this controversy. In an ideal QAH insulator, the quantization of $\rho_{yx}$ and vanishing $\rho_{xx}$ should occur simultaneously, whereas for an AH insulator, a reduced $\rho_{yx}$ and insulating $\rho_{xx}$ are expected. However, when the system just enters the AH insulator regime, such as sample #S13, the $\rho_{yx}$ value is quite close to $h/e^2$ but $\rho_{xx}$ is very large. This behavior is reminiscent of the quantized Hall insulator phase with quantized $\rho_{yx}$ and insulating $\rho_{xx}$ in the QH system [38], which has been found to lie between the QH liquid and Hall insulator phases. Due to the unavoidable inhomogeneities in magnetic TIs, the QAH insulator sample may contain a small fraction of quantized AH insulator phase, which will give rise to a sizable $\rho_{xx}$ and nearly quantized $\rho_{yx}$.

Notably, recent studies in some stoichiometric topological magnets also reveal large but unquantized AH or topological Hall effect [55-60], which arises from enhanced Berry curvature near the Fermi surface similar to magnetically doped TIs. However, the physics for the absence of quantization is different from the AH insulator found here. In the former cases, it is mainly due to the contributions of bulk conduction channels. Whereas in the AH insulator phase, it is attributed to the magnetic-disorder-induced exchange gap fluctuations [8].

Besides the different scaling parameter $\kappa$ and critical resistivity value compared to the QH

phase transition, magnetic field also plays a drastically different role in the QAH phenomena. In QH effect, magnetic field directly couples to itinerant electrons through the orbital effect, which alters the kinetic energy in the form of Landau levels. But for the QAH effect, magnetic field mainly couples with the local moments through the Zeeman energy, which in turn affects the magnetic domains and chiral edge state transport. As a consequence, in the magnetic-field-induced QAH insulator to AH insulator transition the insulating phase exists in the low field side of the QCP, which is opposite to that in the QH liquid to Hall insulator transition. These differences strongly suggest that the QAH phase is not a simple zero-field version of conventional QH effect. We note that $\kappa$ itself is not a universal exponent but is instead expressed as $\kappa \sim 1/vz$, where $v$ is the critical correlation length term and $z$ is the dynamic term. Because both $v$ and $z$ determine the class of quantum criticality for a quantum phase transition, the exact universality class of the QAH phase transition awaits further investigation to disentangle the critical and dynamic exponents.

Although the zero-field version of Hall insulator, the AH insulator, discovered here looks superficially similar to the Anderson insulator reported previously [28], the physics between them are highly distinct in several ascpects. The Anderson insulator phase is induced by the on-site potential disorder [61,62], and the exchange gap is a constant in space. In contrast, the AH insulator phase originates from the magnetic disorder, and the fluctuation of exchange gap plays a dominant role. Because the two types of disorders exhibit different responses to external parameters, it is natural to expect distinct transport signatures in the Anderson insulator and AH insulator phases. For example, electrostatic potential fluatuation usually does not change with temperature, whereas the exchange gap fluctuation due to magnetic disorder decreases with lower temperatures because of better magnetic order. This fundamental distinction may explain the distinct ground state properties of $\rho_{yx}$ between the two phases, both in the vast discrepancy in the absolute values and the opposite temperature dependent behaviors. The QH related edge state physics in the AH insulator phase and the almost quantized $\rho_{yx}$ at the ground state also make it different from the conventional magnetic semiconductors.

In conclusion, we find that magnetic TIs with different degrees of magnetic disorder can be classified into two novel phases denotated as the QAH insulator and AH insualtor phases. In the low-disorder limit of the QAH insulator phase, a universal quantum resistance $h/e^2$ is observed at $H_C$. In the AH insulator samples, magnetic field can drive it to the QAH insulator phase through a quantum phase transition. Both the scaling parameters and the critical resistivity, as well as the role of magnetic field, exhibit apparent discrepancies from that in the QH phase transition. We propose a phenomenological model involving the transmission between chiral edge states at domain boundaries, which explains the novel ground states and peculiar quantum transport phenomena in the QAH effect.

**Figure Captions:**

FIG. 1. Transport properties of five representative magnetic TIs. Magnetic field dependent $\rho_{yx}$ (upper panel) and $\rho_{xx}$ (lower panel) measured at varied $T$s for (a) #S1, (b) #S2, (c) #S3, (d) #S4, and (e) #S5. All the measurements were performed when $V_g$ is tuned to the CNP regime. All the five samples show similar Hall traces, although the $\rho_{yx}^0$ value decreases from 0.96 $h/e^2$ in sample #S1 to 0.56 $h/e^2$ in sample #S5. The $\rho_{xx}$ values increase from sample #S1 to #S5, indicating the increase of disorder. (f)-(g) Extracted $T$ dependent $\rho_{xx}^0$ and $\rho_{xx}^{Hc}$ for the five samples. (h) Evolution of $\rho_{xx}^{Hc}$ as $\rho_{xx}^0$ at different $T$s in the five samples.

FIG. 2. QAH insulator and AH insulator behaviors in more samples. (a)-(b) Magnetic field dependent $\rho_{yx}$ and $\rho_{xx}$ in six QAH insulator samples (#S6 to #S11) and five AH insulator (#S12 to #16) samples. All the data were measured at $T$ = 1.6 K. (c) Experimental phase diagram summarized from the transport data of 16 QAH samples. The error bars result from the uncertainty of the geometrical size of the devices. The blue horizontal broken line marks the universal quantum resistance of $\rho_{xx}^{Hc}$ in the low disorder regime of the QAH insulator phase. The vertical black broken line at $\rho_{xx}^0$ ~ 2 $h/e^2$ marks the phase boundary between the QAH insulator and AH insulator phase.

FIG. 3. Low $T$ transport properties of two representative QAH samples. (a)-(b) Magnetic field dependent $\rho_{xx}$ and $\rho_{yx}$ of a QAH insulator (#S6) and an AH insulator (#S13) sample at low $T$s with $V_g$ tuned to the CNP. (c)-(d) Extracted $T$ dependent data of $\rho_{xx}^0$ (red) and $\rho_{yx}^0$ (blue) for sample #S6 and #S13 respectively. In both samples, $\rho_{yx}^0$ approaches the $h/e^2$ at low $T$, but $\rho_{xx}^0$ exhibits opposite $T$ dependence.

FIG. 4. Magnetic-field-driven quantum phase transition between the QAH insulator and AH insulator states. (a) Magnetic field dependent $\rho_{xx}$ in the AH insulator sample (#S13) measured at varied $T$s. (b) $T$ dependent $\rho_{xx}$ extracted from (a) at different magnetic fields. (c) Scaling analysis of $\rho_{xx}$ in the vicinity of the QCP as a function of $(H\text{-}H_T)/T^{\kappa}$. All the curves collapse together for $\kappa$ ~ 0.31.

FIG. 5. Schematic illustrations of the chiral edge state transmission picture. (a) The hopping transmission between two neighboring chiral edge states in an AH insulator sample at zero field. The blue puddles represent the out-of-plane magnetized domains and the red halos represent the chiral edge states at domain boundaries. The grey region represents area without out-of-plane magnetic order. The lower panel depicts the spatial distribution of chiral edge states. (b) The ohmic transmission in a QAH insulator sample when neighboring magnetic domains are in close proximity to each other. (c) Transmission between two edge states with opposite chirality in a QAH insulator sample at $H_C$, when the neighboring domains have opposite magnetization. The numbers on the edges denote the chemical potential at different sections of the boundaries.

**Data availability:** All raw and derived data used to support the findings of this work are available from the authors on request.

**Acknowledgements:** This work was supported by the Basic Science Center Project of NSFC under grant No. 51788104, MOST of China grant 2017YFA0302900. This work is supported in part by the Beijing Advanced Innovation Center for Future Chip (ICFC).

**Author contributions:** C.L. and Y.W. proposed and designed the research. C.L., Y.F., W.W., S.L., and Z.C. carried out the transport experiments and data analysis. Y.O., G.J., X.M. and K.H. grew the thin films using molecular beam epitaxy. Q.X. supervised the sample growth. Y.W. supervised transport experiments and coordinated the collaborations. C.L. and Y.W. prepared the manuscript with comments from all authors.

## References:

[1] C. Z. Chang *et al.*, *Experimental Observation of the Quantum Anomalous Hall Effect in a Magnetic Topological Insulator*. Science **340**, 167 (2013).

[2] J. G. Checkelsky, R. Yoshimi, A. Tsukazaki, K. S. Takahashi, Y. Kozuka, J. Falson, M. Kawasaki, and Y. Tokura, *Trajectory of the Anomalous Hall Effect Towards the Quantized State in a Ferromagnetic Topological Insulator*. Nat. Phys. **10**, 731 (2014).

[3] X. Kou *et al.*, *Scale-Invariant Quantum Anomalous Hall Effect in Magnetic Topological Insulators Beyond the Two-Dimensional Limit*. Phys. Rev. Lett. **113**, 137201 (2014).

[4] R. Yu, W. Zhang, H. J. Zhang, S. C. Zhang, X. Dai, and Z. Fang, *Quantized Anomalous Hall Effect in Magnetic Topological Insulators*. Science **329**, 61 (2010).

[5] C. X. Liu, X. L. Qi, X. Dai, Z. Fang, and S. C. Zhang, *Quantum Anomalous Hall Effect in $Hg_{1-Y}mn_yte$ Quantum Wells*. Phys. Rev. Lett. **101**, 146802 (2008).

[6] K. Nomura and N. Nagaosa, *Surface-Quantized Anomalous Hall Current and the Magnetoelectric Effect in Magnetically Disordered Topological Insulators*. Phys. Rev. Lett. **106**, 166802 (2011).

[7] E. O. Lachman *et al.*, *Visualization of Superparamagnetic Dynamics in Magnetic Topological Insulators*. Sci. Adv. **1**, e1500740 (2015).

[8] I. Lee *et al.*, *Imaging Dirac-Mass Disorder from Magnetic Dopant Atoms in the Ferromagnetic Topological Insulator $Cr_x(Bi_{0.1}sb_{0.9})_{2-X}te_3$*. Proc. Natl. Acad. Sci. U.S.A. **112**, 1316 (2015).

[9] W. B. Wang, Y. B. Ou, C. Liu, Y. Y. Wang, K. He, Q. K. Xue, and W. D. Wu, *Direct Evidence of Ferromagnetism in a Quantum Anomalous Hall System*. Nat. Phys. **14**, 791 (2018).

[10] Z. Yue and M. E. Raikh, *Smearing of the Quantum Anomalous Hall Effect Due to Statistical Fluctuations of Magnetic Dopants*. Phys. Rev. B. **94**, 155313 (2016).

[11] Y. X. Xing, F. M. Xu, K. T. Cheung, Q. F. Sun, J. Wang, and Y. G. Yao, *Influence of Magnetic Disorders on Quantum Anomalous Hall Effect in Magnetic Topological Insulator Films Beyond the Two-Dimensional Limit*. New J Phys **20**, 043011 (2018).

[12] C. Z. Chen, H. Liu, and X. C. Xie, *Effects of Random Domains on the Zero Hall Plateau in the Quantum Anomalous Hall Effect*. Phys. Rev. Lett. **122**, 026601 (2019).

[13] A. Haim, R. Ilan, and J. Alicea, *Quantum Anomalous Parity Hall Effect in Magnetically Disordered Topological Insulator Films*. Phys. Rev. Lett. **123** (2019).

[14] S. Kudla, A. Dyrdal, V. K. Dugaev, J. Berakdar, and J. Barnas, *Conduction of Surface Electrons in a Topological Insulator with Spatially Random Magnetization*. Phys. Rev. B. **100**, 205428 (2019).

[15] K. L. Tiwari, W. A. Coish, and T. Pereg-Barnea, *Magnetoconductance Signatures of Chiral Domain-Wall Bound States in Magnetic Topological Insulators*. Phys. Rev. B. **96**, 235120 (2017).

[16] Z. Qiao, Y. Han, L. Zhang, K. Wang, X. Deng, H. Jiang, S. A. Yang, J. Wang, and Q. Niu, *Anderson Localization from the Berry-Curvature Interchange in Quantum Anomalous Hall Systems*. Phys. Rev. Lett. **117**, 056802 (2016).

[17] D. A. Abanin and D. A. Pesin, *Ordering of Magnetic Impurities and Tunable Electronic Properties of Topological Insulators*. Phys. Rev. Lett. **106**, 136802 (2011).

[18] A. J. Bestwick, E. J. Fox, X. Kou, L. Pan, K. L. Wang, and D. Goldhaber-Gordon, *Precise Quantization of the Anomalous Hall Effect near Zero Magnetic Field*. Phys. Rev. Lett. **114**, 187201 (2015).

[19] C. Z. Chang *et al.*, *High-Precision Realization of Robust Quantum Anomalous Hall State in a Hard Ferromagnetic Topological Insulator*. Nat. Mater. **14**, 473 (2015).

[20] Y. Ou *et al.*, *Enhancing the Quantum Anomalous Hall Effect by Magnetic Codoping in a Topological Insulator*. Adv. Mater. **30**, 1703062 (2018).

[21] C. Z. Chang, W. Zhao, D. Y. Kim, P. Wei, J. K. Jain, C. Liu, M. H. Chan, and J. S. Moodera, *Zero-Field Dissipationless Chiral Edge Transport and the Nature of Dissipation in the Quantum Anomalous Hall State*. Phys. Rev. Lett. **115**, 057206 (2015).

[22] A. Kandala, A. Richardella, S. Kempinger, C. X. Liu, and N. Samarth, *Giant Anisotropic Magnetoresistance in a Quantum Anomalous Hall Insulator*. Nat. Commun. **6**, 7434 (2015).

[23] Y. Feng *et al.*, *Observation of the Zero Hall Plateau in a Quantum Anomalous Hall Insulator*. Phys. Rev. Lett. **115**, 126801 (2015).

[24] M. Liu, W. Wang, A. R. Richardella, A. Kandala, J. Li, A. Yazdani, N. Samarth, and N. P. Ong, *Large Discrete Jumps Observed in the Transition between Chern States in a Ferromagnetic Topological Insulator*. Sci. Adv. **2**, e1600167 (2016).

[25] M. Mogi, R. Yoshimi, A. Tsukazaki, K. Yasuda, Y. Kozuka, K. S. Takahashi, M. Kawasaki, and Y. Tokura, *Magnetic Modulation Doping in Topological Insulators toward Higher-Temperature Quantum Anomalous Hall Effect*. Appl. Phys. Lett. **107**, 2401 (2015).

[26] S. Grauer, K. M. Fijalkowski, S. Schreyeck, M. Winnerlein, K. Brunner, R. Thomale, C. Gould, and L. W. Molenkamp, *Scaling of the Quantum Anomalous Hall Effect as an Indicator of Axion Electrodynamics*. Phys. Rev. Lett. **118**, 246801 (2017).

[27] X. Feng *et al.*, *Thickness Dependence of the Quantum Anomalous Hall Effect in Magnetic Topological Insulator Films*. Adv Mater **28**, 6386 (2016).

[28] C. Z. Chang, W. Zhao, J. Li, J. K. Jain, C. Liu, J. S. Moodera, and M. H. Chan, *Observation of the Quantum Anomalous Hall Insulator to Anderson Insulator Quantum Phase Transition and Its Scaling Behavior*. Phys. Rev. Lett. **117**, 126802 (2016).

[29] C. Kastl, P. Seifert, X. Y. He, K. H. Wu, Y. Q. Li, and A. Holleitner, *Chemical Potential Fluctuations in Topological Insulator* $(Bi_{0.5}sb_{0.5})_2te_3$ *Films Visualized by Photocurrent Spectroscopy*. 2d Mater **2** (2015).

[30] Q. L. He *et al.*, *Chiral Majorana Fermion Modes in a Quantum Anomalous Hall Insulator-Superconductor Structure*. Science **357**, 294 (2017).

[31] W. J. Ji and X. G. Wen, *1/2 ($E^2$/H) Conductance Plateau without 1d Chiral Majorana Fermions*. Phys. Rev. Lett. **120**, 107002 (2018).

[32] Y. Y. Huang, F. Setiawan, and J. D. Sau, *Disorder-Induced Half-Integer Quantized Conductance Plateau in Quantum Anomalous Hall Insulator-Superconductor Structures*. Phys. Rev. B. **97**, 107002 (2018).

[33] M. Kayyalha *et al.*, *Absence of Evidence for Chiral Majorana Modes in Quantum Anomalous Hall-*

*Superconductor Devices*. Science **367**, 64 (2020).

[34] S. Kivelson, D. H. Lee, and S. C. Zhang, *Global Phase Diagram in the Quantum Hall Effect*. Phys. Rev. B. **46**, 2223 (1992).

[35] D. Shahar, D. C. Tsui, M. Shayegan, R. N. Bhatt, and J. E. Cunningham, *Universal Conductivity at the Quantum Hall Liquid to Insulator Transition*. Phys. Rev. Lett. **74**, 4511 (1995).

[36] L. W. Wong, H. W. Jiang, N. Trivedi, and E. Palm, *Disorder-Tuned Transition between a Quantum Hall Liquid and Hall Insulator*. Phys. Rev. B. **51**, 18033 (1995).

[37] W. Pan, D. Shahar, D. C. Tsui, H. P. Wei, and M. Razeghi, *Quantum Hall Liquid-to-Insulator Transition in $In_{1-X}ga_{x}as$/Inp Heterostructures*. Phys. Rev. B. **55**, 15431 (1997).

[38] M. Hilke, D. Shahar, S. H. Song, D. C. Tsui, Y. H. Xie, and D. Monroe, *Experimental Evidence for a Two-Dimensional Quantized Hall Insulator*. Nature **395**, 675 (1998).

[39] A. F. Hebard and M. A. Paalanen, *Magnetic-Field-Tuned Superconductor-Insulator Transition in 2-Dimensional Films*. Phys. Rev. Lett. **65**, 927 (1990).

[40] A. Yazdani and A. Kapitulnik, *Superconducting-Insulating Transition in 2-Dimensional Alpha-Moge Thin-Films*. Phys. Rev. Lett. **74**, 3037 (1995).

[41] N. Nagaosa, J. Sinova, S. Onoda, A. H. MacDonald, and N. P. Ong, *Anomalous Hall Effect*. Rev. Mod. Phys. **82**, 1539 (2010).

[42] A. M. Pruisken, *Universal Singularities in the Integral Quantum Hall Effect*. Phys. Rev. Lett. **61**, 1297 (1988).

[43] H. P. Wei, L. W. Engel, and D. C. Tsui, *Current Scaling in the Integer Quantum Hall Effect*. Phys. Rev. B. **50**, 14609 (1994).

[44] A. J. M. Giesbers, U. Zeitler, L. A. Ponomarenko, R. Yang, K. S. Novoselov, A. K. Geim, and J. C. Maan, *Scaling of the Quantum Hall Plateau-Plateau Transition in Graphene*. Phys. Rev. B. **80**, 241411 (R) (2009).

[45] W. Li, C. L. Vicente, J. S. Xia, W. Pan, D. C. Tsui, L. N. Pfeiffer, and K. W. West, *Scaling in Plateau-to-Plateau Transition: A Direct Connection of Quantum Hall Systems with the Anderson Localization Model*. Phys. Rev. Lett. **102**, 216801 (2009).

[46] E. Shimshoni, S. L. Sondhi, and D. Shahar, *Duality near Quantum Hall Transitions*. Phys. Rev. B. **55**, 13730 (1997).

[47] X. Leng, J. Garcia-Barriocanal, S. Bose, Y. Lee, and A. M. Goldman, *Electrostatic Control of the Evolution from a Superconducting Phase to an Insulating Phase in Ultrathin $Yba_2cacu_3o_{7-X}$ Films*. Phys. Rev. Lett. **107** (2011).

[48] Y. Huo, R. E. Hetzel, and R. N. Bhatt, *Universal Conductance in the Lowest Landau Level*. Phys.

Rev. Lett. **70**, 481 (1993).

[49] E. Shimshoni and A. Auerbach, *Quantized Hall Insulator: Transverse and Longitudinal Transport*. Phys. Rev. B. **55**, 9817 (1997).

[50] L. P. Pryadko and A. Auerbach, *Hall Resistivity and Dephasing in the Quantum Hall Insulator*. Phys. Rev. Lett. **82**, 1253 (1999).

[51] U. Zulicke and E. Shimshoni, *Quantum Breakdown of the Quantized Hall Insulator*. Phys. Rev. B. **63**, 241301 (R) (2001).

[52] M. Allen, Y. T. Cui, E. Y. Ma, M. Mogi, M. Kawamura, I. C. Fulga, D. Goldhaber-Gordon, Y. Tokura, and Z. X. Shen, *Visualization of an Axion Insulating State at the Transition between 2 Chiral Quantum Anomalous Hall States*. Proc. Natl. Acad. Sci. U.S.A. **116**, 14511 (2019).

[53] K. Yasuda, M. Mogi, R. Yoshimi, A. Tsukazaki, K. S. Takahashi, M. Kawasaki, F. Kagawa, and Y. Tokura, *Quantized Chiral Edge Conduction on Domain Walls of a Magnetic Topological Insulator*. Science **358**, 1311 (2017).

[54] J. Wang, B. Lian, H. J. Zhang, and S. C. Zhang, *Anomalous Edge Transport in the Quantum Anomalous Hall State*. Phys. Rev. Lett. **111**, 086803 (2013).

[55] S. Nakatsuji, N. Kiyohara, and T. Higo, *Large Anomalous Hall Effect in a Non-Collinear Antiferromagnet at Room Temperature*. Nature **527**, 212 (2015).

[56] X. K. Li *et al.*, *Chiral Domain Walls of* $Mn_3sn$ *and Their Memory*. Nat. Commun. **10**, 3021 (2019).

[57] J. M. Taylor, A. Markou, E. Lesne, P. K. Sivakumar, C. Luo, F. Radu, P. Werner, C. Felser, and S. S. P. Parkin, *Anomalous and Topological Hall Effects in Epitaxial Thin Films of the Noncollinear Antiferromagnet* $Mn_3sn$. Phys. Rev. B. **101**, 094404 (2020).

[58] L. Ye *et al.*, *Massive Dirac Fermions in a Ferromagnetic Kagome Metal*. Nature **555**, 638 (2018).

[59] C. Liu, Y. Zang, W. Ruan, Y. Gong, K. He, X. Ma, Q. K. Xue, and Y. Wang, *Dimensional Crossover-Induced Topological Hall Effect in a Magnetic Topological Insulator*. Phys. Rev. Lett. **119**, 176809 (2017).

[60] J. Jiang *et al.*, *Concurrence of Quantum Anomalous Hall and Topological Hall Effects in Magnetic Topological Insulator Sandwich Heterostructures*. Nat Mater **19**, 732 (2020).

[61] C. W. Groth, M. Wimmer, A. R. Akhmerov, J. Tworzydlo, and C. W. Beenakker, *Theory of the Topological Anderson Insulator*. Phys. Rev. Lett. **103**, 196805 (2009).

[62] J. Li, R. L. Chu, J. K. Jain, and S. Q. Shen, *Topological Anderson Insulator*. Phys. Rev. Lett. **102**, 136806 (2009).

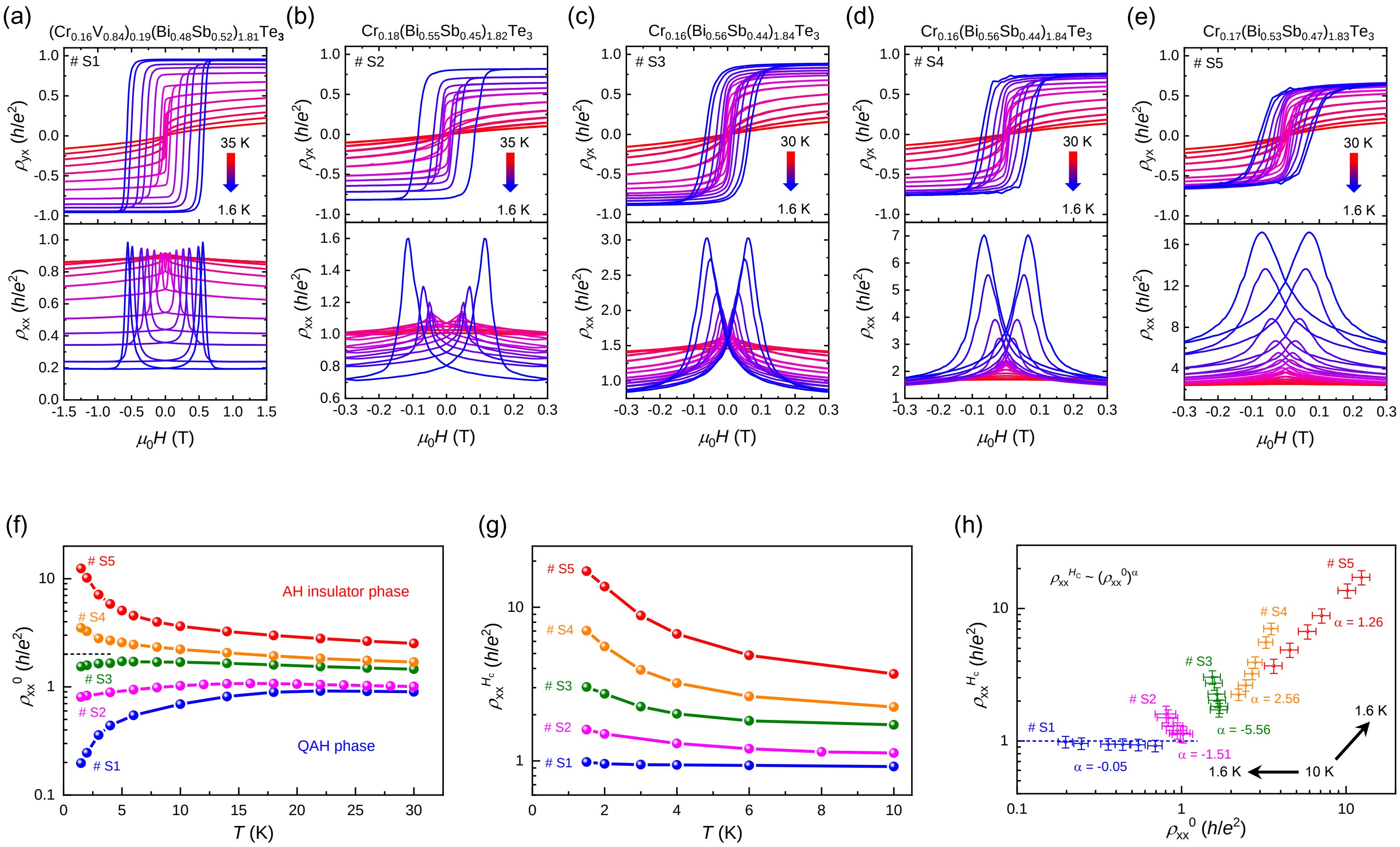

Figure 1

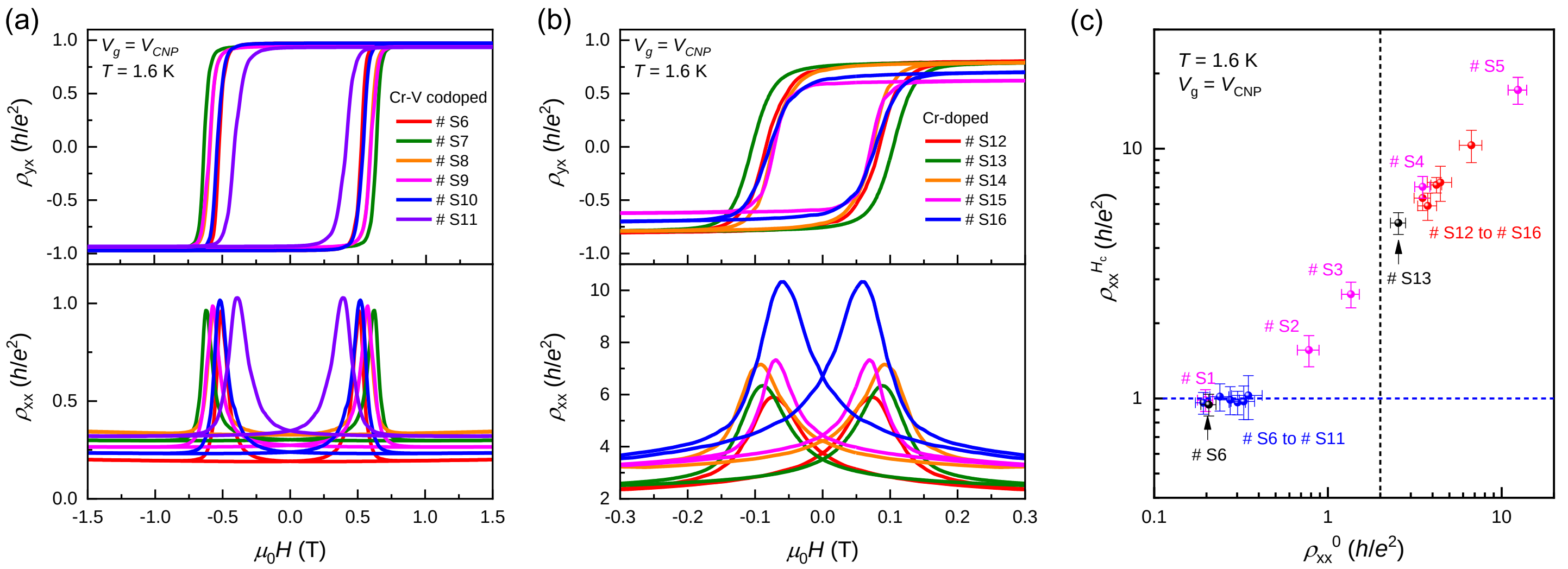

Figure 2

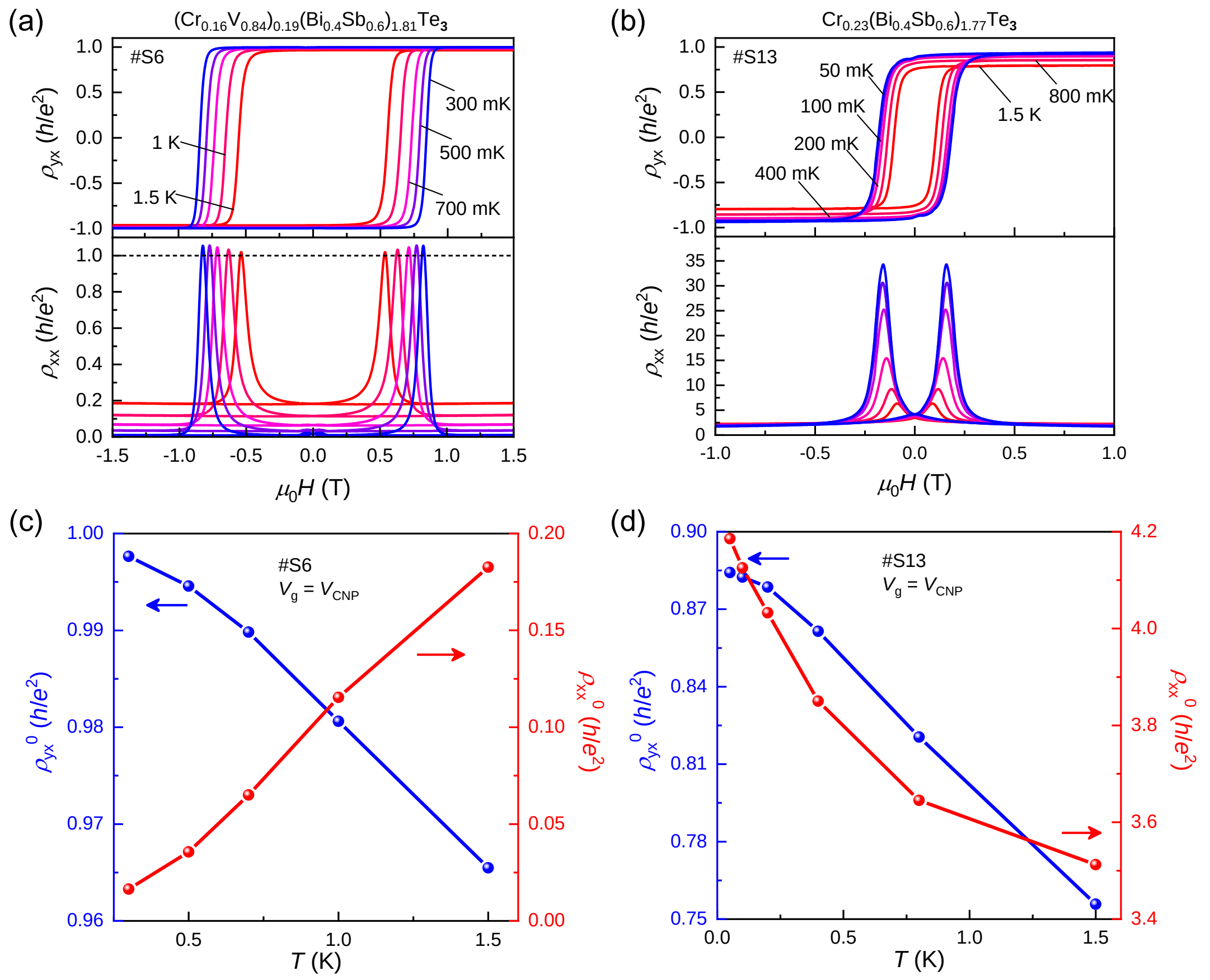

Figure 3

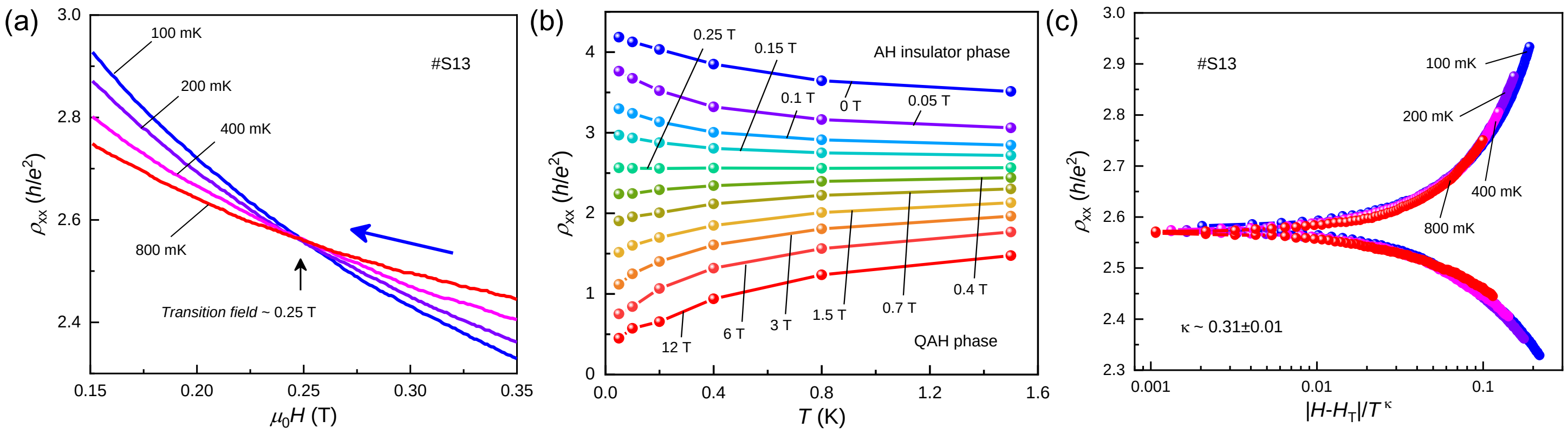

Figure 4

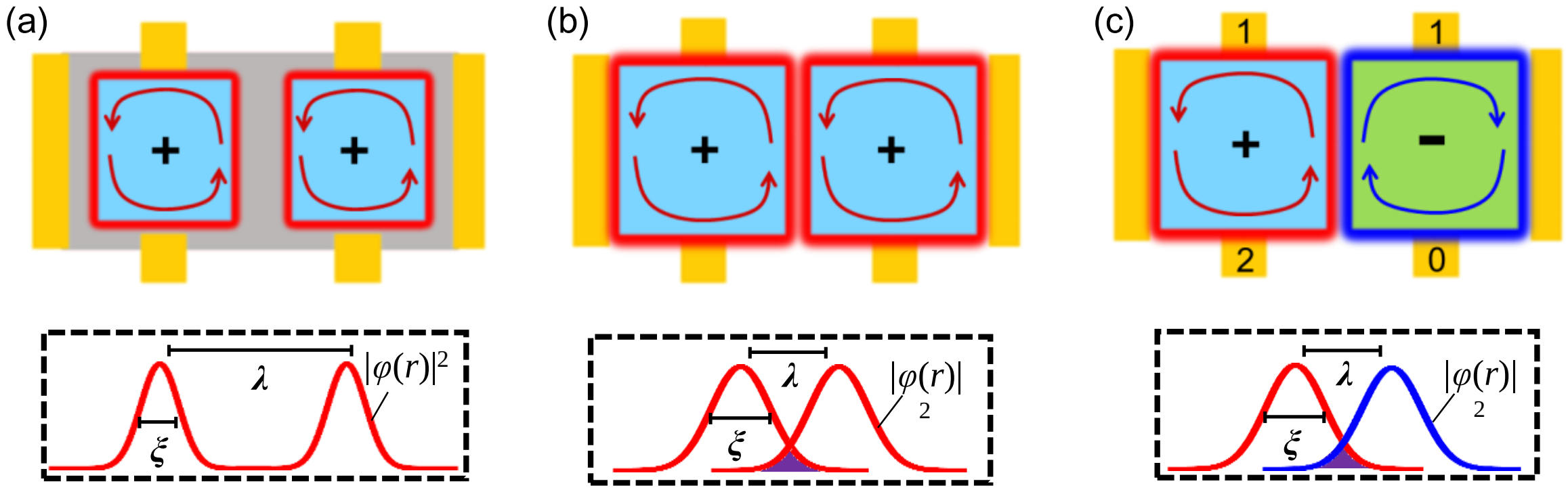

Figure 5